\documentstyle[aps,epsfig]{revtex}
\draft 
\begin{document}
\title{Interferometry with Resonances and Flow in High-Energy Nuclear 
Collisions}
\author{H. Heiselberg}
\address{NORDITA, Blegdamsvej 17, DK-2100 Copenhagen \O., Denmark}
\date{Feb., 1996}
\maketitle

\begin{abstract}
The effects of resonances and flow on the correlation function for two
identical particles are described assuming chaotic
sources and classical propagation of particles.  Expanding to second
order in relative momenta, the source sizes can be calculated directly
and understood as contributions from various fluctuations in the
source.  Specific calculations of source size radii are given assuming
Bjorken longitudinal flow with additional transverse expansion.
Results are compared to recent $\pi\pi$ and $KK$ correlation data from
relativistic nuclear collisions with particular attention to
the reduction in the $\pi\pi$ correlation function 
due to resonances and the decreasing source sizes with
increasing transverse momenta of the particles.
\end{abstract}

\noindent

\section{Introduction} 

 The HBT effect based on interference of identical particles
\cite{HBT,pp} is an important method in relativistic heavy ion
collisions for extracting information about the source sizes and life
times. \cite{bengt} The correlations due to interference are well
described for {\it pp} collisions \cite{ppL} when the effect of
resonances in particular the long lived ones are included.  In high
energy nuclear collisions we hope to extract the spatial, temporal and
momentum distribution of particles in the source at decoupling or
freeze-out by correcting for resonances, Coulomb effects, and possible
other final state interactions. The source sizes, life times and flow
effects and their dependence on collision energy, impact parameter,
projectile and target mass will be crucial for determining whether a
quark-gluon plasma is created.  Recent data allow determination of
longitudinal, outwards and sidewards source radii for pions and kaons
in heavy ion collisions at CERN and Brookhaven energies.  The rapidity
and transverse momentum dependence of source sizes have also been
measured. \cite{NA44,NA35} The rapidity dependence of the longitudinal
source size agrees with longitudinal expansion whereas decreasing
transverse radii seem to indicate that also transverse expansion takes
place \cite{Csorgo,Heinz}.

Resonances are known to be abundant in relativistic heavy ion
collisions and they contribute significantly to pion production by,
e.g., $\sim$80\% according to the Fritiof model. \cite{Fritiof}
Resonances affect the source such that it seems to have a larger life
time or larger outward source size \cite{GP}. The very long lived
resonances are not resolved and reduce correlations. 
However, the reduction is also expected if the
source is partially coherent.

The purpose of this paper is to do a combined analysis of resonances
and flow both longitudinal and transverse, and to calculate the
nontrivial interplay. For that purpose the the correlation function
for an incoherent source with plane wave propagation is extended to
include resonances. By expanding in small relative momenta between the
identical particles, one can extract the source radii analytically for
general cylindrical symmetric sources with longitudinal and transverse
flow. The effect of resonances and flow on the correlation function
will be discussed and compared to recent NA44 data, which can extract
sidewards, outwards and longitudinal source sizes.

\section{Source size fluctuations}

Assuming an incoherent source and plane wave propagation, 
the correlation function reduces to the simple Fourier transform 
\cite{Pratt,GKW}
\begin{eqnarray}
   C({\bf q},{\bf K}) = 1\pm \frac{|\int d^4x \, S(x,K)e^{iq\cdot x}|^2}
                        {|\int d^4x \, S(x,K)|^2} \, , \label{C1}
\end{eqnarray}
for two identical bosons/fermions ($\pm$) of 
relative and total momenta $q=k_1-k_2$ and $K=k_1+k_2$ respectively,
when $q\ll K$. The source distribution $S(x,K)=dN/d^4xd{\bf K}$ 
describes the particle production at space-time 
point $x={\rm (t,x,y,z)}$ of a particle with momentum $k_1\simeq k_2\simeq K$.

Resonances will here be included by assuming classical propagation.
The additional distance travelled by the resonance,
$\Delta x=u_r\Delta\tau$, where $\Delta\tau$ is
the life time of the resonance and $u_r$ its velocity, 
gives an extra phase, $\Delta x\cdot q$, in Eq. (\ref{C1}).
When the life times are exponentially distributed with
decay time $\tau_r=1/\Gamma_r$, the correlation function
becomes (see also \cite{GP})
\begin{eqnarray}
   C({\bf q},{\bf K}) &=& 1\pm \frac{|\sum_r \int d^4x \, 
             S_r(x,{\bf K})e^{iq\cdot x}(1-iq\cdot u_r\tau_r)^{-1}|^2}
             {|\sum_r \int d^4x \, S_r(x,{\bf K})|^2}  \, .  \label{C2}
\end{eqnarray}
Now, $x$ refers to the space-time production point of the resonance.

It is convenient notation to introduce the space-time integration 
and summation over resonances of an operator $\tilde{O}$ by the average
\begin{eqnarray}
 \langle \tilde{O}\rangle &\equiv& \frac{|\sum_r \int d^4x \,  S_r(x,{\bf K})
                         \, \tilde{O}|}
                        {|\sum_r \int d^4x S_r(x,{\bf K})|} \, .\label{C3}
\end{eqnarray}
Likewise the fluctuation of that operator is
\begin{eqnarray}
   \sigma(\tilde{O}) =\langle\tilde{O}^2\rangle-\langle\tilde{O}\rangle^2 \,.
\end{eqnarray}
The correlation function for bosons is then
\begin{eqnarray}
   C({\bf q},{\bf K}) &=& 1 \pm \langle 
            e^{iq\cdot x}(1-iq\cdot u_r\tau_r)^{-1}\rangle^2 \, .\label{C4}
\end{eqnarray}
Whereas the full $q$-dependence of the correlation function is generally
quite cumbersome to calculate, the small $q$-dependence can be
calculated directly by expanding (\ref{C4}) to second order in $q$
\begin{eqnarray}  
  C({\bf q},{\bf K}) &=& 1\pm \left[1-\sigma(q\cdot[x+u_r\tau_r])
                       - \langle [q\cdot u_r\tau_r]^2 \rangle 
                       - {\cal O}((q\cdot x)^4)\right] \, .  \label{C5}
\end{eqnarray}
The expansion is valid when $R_\mu q_\mu\ll 1$, where $R_\mu$ are the
space-time extension of the system. If the source and therefore also
the correlation function is approximately gaussian, then
the leading term is sufficient to determine its shape.
Notice that the resonance life-times contributes to the 
correlation function in Eq. (\ref{C5}) through the
fluctuations in space and time, $\sigma(q\cdot[x+u_r\tau_r])$, as well as
contributing by the explicit term $\langle [q\cdot u_r\tau_r]^2 \rangle$.

Experimentalist often parametrize their pion and kaon correlation function by
\begin{eqnarray}
  C({\bf q},{\bf K}) = 1+\lambda \exp[-q_s^2R_s^2-q_o^2R_o^2-q_l^2R_l^2
                       -2q_oq_lR_{ol}^2] \ , \label{Cexp}
\end{eqnarray}
where ${\bf q}_{o,s,l}$ are the usual outwards, sidewards and longitudinal
projections respectively of the relative momentum in a cartesian coordinate 
system where ${\bf q}_l$ lies along the beam (or longitudinal) axis and 
${\bf q}_s$ is perpendicular to both the beam axis and ${\bf K}$. The
cross term with coefficient $R_{ol}$ has recently been advocated by
Heinz et al. \cite{Heinz}.  Indications of this kind of asymmetry
have been seen in experiments \cite{NA35}. The prefactor $\lambda$ is
found to be approximately unity for kaons and about half for pions in most 
experiments. The reduced pion correlations may be due to ``coherence'' 
in the source or to long lived resonances as will be discussed below.

Because the momentum of the pair ${\bf K}$ breaks cylindrical
symmetry and, with the beam line, determines the {\rm x-z} plane, 
the average values for the 
${\rm z}$ and ${\rm x}$ coordinates are generally non-vanishing
in a longitudinally and transversally expanding system. Reflection
symmetry in the ${\rm y}$ coordinate leads to $\langle {\rm y}\rangle=0$ and
we can assume that the resonance velocity $u_r=(u_t,u_x,u_y,u_z)$ 
has vanishing {\rm y}-component, $\langle u_y\rangle=0$. Consequently,
all cross terms with $R_s$ vanish as in (\ref{Cexp}). 
Since $q=(\beta_oq_o+\beta_lq_l,q_o,q_s,q_l)$, where $\beta_i=K_i/K_0$, we
obtain from (\ref{C5}) and (\ref{Cexp}) 
\begin{eqnarray}
   R_s^2 &=& \sigma({\rm y})  \label{Rs}\\
   R_o^2 &=& \sigma({\rm x}+u_x\tau_r-\beta_o(t+u_t\tau_r))
          + \langle[(u_x-\beta_ou_t)\tau_r]^2\rangle  \\
   R_l^2 &=& \sigma({\rm z}+u_z\tau_r-\beta_l(t+u_t\tau_r))
          + \langle[(u_z-\beta_lu_t)\tau_r]^2\rangle \\
   R_{ol}^2 &=& \langle [x+u_x\tau_r-\beta_o(t+u_t\tau_r)]
                [{\rm z}+u_z\tau_r-\beta_l(t+u_t\tau_r)]\rangle \nonumber\\
            &-& \langle x+u_x\tau_r-\beta_o(t+u_t\tau_r)\rangle
                \langle {\rm z}+u_z\tau_r-\beta_l(t+u_t\tau_r)\rangle 
             +  \langle(u_x-\beta_ou_t)(u_z-\beta_lu_t)\tau_r^2\rangle
             \, . \label{Rol}
\end{eqnarray}
 
To evaluate the source radii, we need to specify the source further.
Since the particle spectra have approximately thermal transverse momentum
distributions in the energy and $p_\perp$ regions considered here,
we will here assume that 
all resonances are thermally distributed locally
\begin{eqnarray}
  S_r(x,K) = f_r\exp[-K\cdot u(x)/T_r]\, \rho(x)\, .\label{ST}
\end{eqnarray}
Here, $f_r$ determines the fraction of particles coming from resonance $r$,
$T_r$ is the ``temperature'' or more accurately the 
inverse of the $m_\perp$ slope of particles produced through
that resonance, $u(x)$ is the flow velocity at a given point 
in space and time, and $\rho$ is the spatial and temporal source function.
For longitudinally expanding systems the proper time
$\tau=\sqrt{t^2-{\rm z}^2}$ and 
$\eta=\frac{1}{2}\ln((t+{\rm z})/(t-{\rm z}))$ are convenient variables 
whereby $d^4x=\tau d\tau d\eta d{\rm x} d{\rm y}$.

The source function is commonly parametrized
by factorizing gaussians common for all resonances 
\begin{eqnarray}
   \rho(\tau,\eta,{\rm x,y}) \sim \frac{1}{\tau}\exp\left[
    -\frac{{\rm x^2+y^2}}{2\sigma_\perp}-\frac{(\eta-\eta_0)^2}{2\sigma_\eta}
       -\frac{(\tau-\langle\tau\rangle)^2}{2\sigma_\tau}\right] \, . \label{Rg}
\end{eqnarray}
However, the detailed form will not be needed in the present
analysis. Only the mean and the fluctuations in the quantities, e.g.,
$\langle {\rm x}\rangle$ and $\sigma({\rm x})=\langle {\rm x}^2\rangle
-\langle {\rm x}\rangle^2$, are needed to leading order. When the
experimental accurracy improves one may be able to measure finer
details of correlation functions which then can determine higher
moments.  Whereas we can always translate our coordinate system so
that $\langle {\rm x}\rangle=\langle{\rm
y}\rangle=\langle\eta\rangle=0$, the average emission or freeze-out
time $\langle\tau\rangle$ does not vanish and it determines the
longitudinal extension of the source.  The source in Eq. (\ref{Rg})
assumes cylindrical symmetry, i.e. $\sigma({\rm x})=\sigma({\rm y})$.
If one is able to determine the plane of reaction, that symmetry is
broken \cite{AGS} and one may determine $\sigma({\rm x})$ and
$\sigma({\rm y})$ separately.

\section{Longitudinal expanding system}

Let us first analyse the situation with longitudinal expansion as in the
Bjorken model, i.e., $u=(\cosh\eta,0,0,\sinh\eta)$. 
Defining the mean rapidity as $Y=(y_1+y_2)/2$, we find
\begin{eqnarray}
   K\cdot u = m_\perp\cosh(\eta-Y) \, .
\end{eqnarray}
The resonance velocities are approximated by their local average over
all directions which is simply the local flow velocity, i.e.,
$u_r=u$. 
It is now straight forward to evaluate the source radii of 
(\ref{Rs}-\ref{Rol}). However,
the gaussian in rapidity has the effect of moving the average
rapidity, $\langle\eta\rangle$ (referred to as the ``saddle point'' in Refs. 
\cite{Heinz,Csorgo}), away from the mean rapidity, $Y$, of the
two interfering particles. 
In the limit when $\sigma_\eta\gg T/m_\perp$ so that
$\langle\eta\rangle\simeq Y$, the results simplify to
\begin{eqnarray}
   R_s^2 &=& \sigma({\rm y})  \label{Rsb} \\
   R_o^2 &=& \sigma({\rm x}) +\beta_\perp \left[\sigma(\tau)
         +\sigma(\tau_r)+ \langle\tau_r^2\rangle
         +\langle\tau+\tau_r\rangle^2 \tanh^2Y\sigma(\eta)\right] \label{Rob}\\
   R_l^2 &=& \sigma(\eta) \langle\tau^2+2\tau\tau_r
              +2\tau_r^2\rangle \cosh^{-2}(Y)  \label{Rl} \\
   R_{ol}^2 &=& - \beta_\perp \sigma(\eta)\langle\tau+\tau_r\rangle^2 
                \sinh(Y)\cosh^{-2}(Y)  \label{Rolb} \, ,
\end{eqnarray}
where $\beta_\perp=p_\perp/m_\perp$.  We observe that the source sizes
are given in terms of the dispersion of the source, namely the
fluctuations in spatial coordinates ($\sigma({\rm x})$ and
$\sigma({\rm y})$), temporally ($\sigma(\tau)$), rapidity
($\sigma(\eta)$), and in resonance life times ($\sigma(\tau_r)$).  For
the simplified source of Eq. (\ref{Rg}) the spatial fluctuations
transversally are simply $\sigma({\rm x})=\sigma({\rm
y})=\sigma_\perp$, and the temporal fluctuations are
$\sigma(\tau)=\sigma_\tau$.

The fluctuation in rapidity consist not only of the spatial rapidity 
fluctuation in the distribution (\ref{Rg}) but is strongly reduced by the
additional rapidity dependence present in the thermal factor of (\ref{ST})
\begin{eqnarray}
  \frac{1}{\sigma(\eta)} = \langle\frac{m_\perp}{T_r}\rangle 
               +\frac{1}{\sigma_\eta}     \, .\label{seta}
\end{eqnarray}
However, as discussed above, it was assumed that $\sigma_\eta\gg T/m_\perp$
in deriving Eqs. (\ref{Rsb}-\ref{Rolb}) and therefore the first term in 
(\ref{seta}) dominates. 
When the various resonances have different temperatures,
the average over $T_r$ will produce fluctuations in temperature, 
$\sigma(T_r)$, besides the average value $\langle T_r\rangle$.


\begin{table}
\caption{Resonances, that decay into pions and contribute to the pion 
correlation
function. The fractions are predicted in the Fritiof \protect{\cite{Fritiof}}
and RQMD \protect{\cite{RQMD}} models for central S+Pb collisions
at energies around 200 GeV/A, midrapidity and all $m_\perp$. }
\begin{tabular}{||c|c|c|c|c|c|c||}  
 Resonance   & Decay      &$m_r$  & Width   &$c\tau_r$ &$f_r^{Fritiof}$&
$f_r^{RQMD}$     \\  \tableline
 direct $\pi$&              & 140 MeV &  -      &  0  fm   & 0.19   & 0.33   \\
 $\rho$      &$\pi\pi$      & 770 MeV & 153 MeV & 1.3 fm   & 0.40   & 0.26   \\
 $\Delta$    &$N\pi$        &1232 MeV & 115 MeV & 1.7 fm   & 0.06   & 0.12   \\
 $K^*$       &$K\pi$        & 892 MeV &  50 MeV & 3.9 fm   & 0.09   & 0.07   \\
 $\Sigma^*$  &$\Sigma\pi$   &1385 MeV &  36 MeV & 5.5 fm   & 0.01   & 0.02   \\
 $\omega$    &$\pi\pi\pi$   & 783 MeV & 8.4 MeV & 23.4 fm  & 0.16   & 0.07   \\
 $\eta'$     &$\eta\pi\pi$  & 958 MeV & 0.24MeV & 821 fm   & 0.02   & 0.02   \\
 $\eta$      &$\pi\pi\pi$   & 549 MeV & 1.1 keV & 1.2 \AA  & 0.04   & 0.03   \\
 $K^0_S$     &$\pi\pi$      & 498 MeV &$\sim$ 0 & 2.7 cm   & 0.03   & 0.07   \\
 $\Sigma,\bar{\Sigma}$  &$n\pi,\bar{n}\pi$ 
                            &1193 MeV &$\sim$ 0 & 4.4 cm   & 0.00   & 0.01   \\
\end{tabular}
\end{table}

The resonances contribute to fluctuations by $\sigma(\tau_r)$.
When the temperatures or $m_\perp$-slopes are the same for all resonances,
the fluctuation in the resonance life times is
\begin{eqnarray}
   \sigma(\tau_r) = \langle\tau_r^2\rangle -\langle\tau_r\rangle^2 
                  = \sum_r f_r\tau_r^2 -(\sum_r f_r\tau_r)^2 \, ,\label{stau}
\end{eqnarray}
where $f_r$ is the fraction of pions arising from resonance $r$ as, for
example, given in Table I for the Fritiof \cite{Fritiof} and RQMD 
\cite{RQMD} models. The
resonance contributions may vary significantly with rapidity and transverse
mass, i.e. the $T_r$'s are different, and therefore the 
fraction at the relevant rapidity and transverse mass of the pair, 
$f_r(Y,m_\perp)$, should be inserted in Eq. (\ref{stau}).

In relativistic heavy ion collisions numerous resonances contribute to
pion production as illustrated in Table I.  The relative contributions
from the various resonances have unfortunately not been measured very
accurately and different models give a variety of results. This is
illustrated by the Fritiof and RQMD models in Table I, where the
feed-down to $\pi^+$ around mid-rapidity are given for central $S+Pb$
collision at 200 GeV/A corrected for detection efficiency.  Note that
the rescatterings in RQMD lead to a strangeness enhancement, which at
mid-rapidities is mainly felt as an enhancement of the long-lived
$K^0_S$ and to less degree of hyperons.  Feed-down from
$\Xi^*,\bar{\Lambda},\phi,...$ contribute by a few per mille each and
have not been included in Table I.  The feed-down to kaons is mainly
from the $K^*$ by $\sim$50\% in Fritiof and $\sim$5\% in a thermal
model \cite{Bolz}).

Since the typical minimum relative momentum,
that can be measured between pions in relativistic heavy ion
collisions, is of order $q_{min}\sim 5-10$MeV, the resonance phase is large,
$q\cdot\tau_r\gg1$, for the long lived resonances $r=\eta,\eta',K^0_S,\Sigma,
\bar{\Sigma},...,$
and therefore the small $q$ employed in Eq. (\ref{C5})
expansion is not allowed. Instead the resonance factor vanishes, $|1-iq\cdot
u_r\tau_r|^{-1}\ll 1$, for these long lived resonances
and their contribution to the correlation function can be
neglected.  This can be taken into account by simply excluding these
long lived resonances in the sum over resonances and the
correlation function in (\ref{Rg}) is reduced by a factor
\begin{eqnarray}
   \lambda_{res} = (1-f_\eta-f_{\eta'}-f_{K^0_S}-f_{\Sigma,\bar{\Sigma},..})^2
                     \, .
\end{eqnarray}
The $\omega$ resonance has a life-time such that $q_{min}
\tau_\omega\sim 1$ and should therefore be treated as an intermediate
case. 


\begin{figure}
\centerline{\psfig{figure=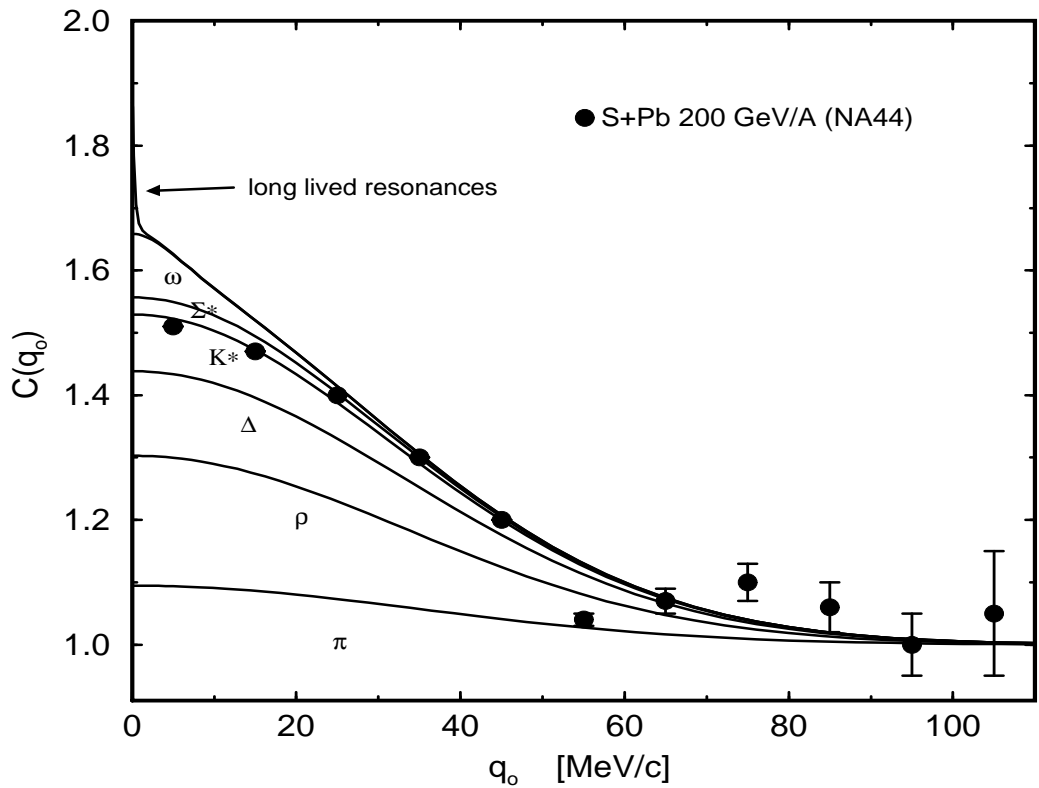,width=16cm,height=12cm,angle=-90}}
\vspace{5mm}
\caption{Resonance contributions 
to the $\pi\pi$ correlation function as function of outwards relative
momentum, $q_o$ (see text).  Curves include successively direct pions,
$\rho$, $\Delta$, $K^*$, $\Sigma^*$, $\omega$ and $\eta+\eta'+K^0_S+\Sigma+
\bar{\Sigma}+...$ 
from RQMD (Table I).  Experimental data for central S+Pb collisions at 
200 GeV/A
\protect{\cite{NA44}} is shown for comparison.}
\label{fig1}
\end{figure}

To illustrate the effects of resonances, the two-pion correlation function
is shown in Fig. 1 as function of the out-ward relative momentum.  The
correlation functions are calculated from Eq. (\ref{C4}) with the
gaussian source of Eq. (\ref{Rg}), which Fourier transforms to
$|\tilde{\rho}|^2=\exp[-q_o^2\sigma_\perp]$. The transverse radius 
$\sqrt{\sigma_\perp}=4$fm is assumed and the resonance fractions and
life times are from RQMD as given in Table I.  The curves shows the
contribution from resonances by adding them successively in the order
of increasing life-times.  It is assumed that the source is
completely incoherent and that Coulomb effects have been corrected
for. The long lived resonances lead to a smaller
$\lambda$ value and the resonances in particular the $\omega$ 
makes the resulting
correlation function steeper than a simple gaussian at small
$q_o$. Many experiments actually find that the pion correlation data
is fit better by steeper functions (e.g., exponentials) than gaussians. 

For comparison, the $\pi^+\pi^+$ correlation function measured in 
S+Pb collisions
at 200 GeV/A \cite{NA44} is also shown in Fig. 1.  The data are
collected for $q_s,q_l\le q_{cut}=20$MeV/c \cite{NA44}. The cuts lead to
a further reduction
\begin{eqnarray}
  \lambda\equiv C(q_o,q_s\le q_{cut},q_l\le q_{cut})-1 
           =    \lambda_{res}\lambda_{cut} \, ,
\end{eqnarray}
where $\lambda_{cut}=(1-R_s^2q_{cut}^2/3)(1-R_l^2q_{cut}^2/3)$ when
$q_{cut}R_{s,l}\raisebox{-.5ex}{$\stackrel{<}{\sim}$} 1$, 
and the quadratic expansion of Eq. (\ref{C5})
is valid. For 
the NA44 source sizes $R_s=4.2$fm and $R_l=4.7$fm and $q_{cut}=20$MeV/c
one finds $\lambda_{cut}=0.87$. 
In the comparison with
experimental data of Fig. 1, the theoretical curves include the
13\% reduction from $\lambda_{cut}$. The data prefers more long
lived resonances than predicted by 
Fritiof (see also \cite{SPACER}) and RQMD
and there is no indication 
of the $\omega$ resonances. Due to strangeness 
enhancement of in particular $K^0_S$, the RQMD model describes the
data better than Fritiof. 
In the recent Pb+Pb
experiments at CERN, the small $\lambda^{\pi\pi}\simeq 0.3$ measured
\cite{NA49} may be explained by the larger $q_{cut}=30$MeV and larger
source sizes $R_{o,s,l}\simeq 6-7$fm, which lead to a smaller
$\lambda_{cut}$. No $\lambda_{cut}$ applies when invariant relative
momenta, $q_{inv}=\sqrt{q^2}$, are employed which explains why
$\lambda_{inv}=C(q_{inv}=0)-1$ generally are found to be larger.

The measured values for the sidewards and outwards radii are very
similar in high energy nuclear collisions. For example, in {\it S+Pb}
collisions \cite{NA44} $R_o=(4.02\pm0.14)$fm and
$R_s=(4.15\pm0.27)$fm so that $R_s^2-R_s^2=(-1.1\pm3.4)$fm$^2$. In
Eq. (\ref{Rob}) the resonances makes a difference of
$\sigma(\tau_r)=1.2$fm$^2$ and $\langle\tau_r^2\rangle=3.0$fm$^2$ in
the Fritiof model even without including $\omega$'s or the long lived
resonances. The transverse velocities for the measured pions are
$\beta_\perp\sim 0.6-0.9$. Thus the short lived resonances lead to a
significant difference between $R_o^2-R_s^2$ that is already between one and
two standard deviations larger than the experimental value. This leaves very
little room for additional fluctuations in source emission time, 
$\sigma(\tau)$, and therefore the pions must be produced in a ``flash''.

Due to longitudinal expansion, both the expansion time of the source
$\langle\tau\rangle$ and the lifetime of the resonances $\tau_r$ as
well as their fluctuations contribute to $R_o$, $R_l$ and $R_{ol}$.
The longitudinal source size scales with rapidity and transverse mass
as $R_l\propto (\sqrt{m_\perp}\cosh(Y))^{-1}$ as found in \cite{AS}
and according to Eq. (\ref{Rl}) when $\sigma_\eta\ll
m_\perp/T$. Experiments confirm both the rapidity dependence
\cite{NA35,NA49} and the $m_\perp$ dependence (see \cite{NA44} and
Fig. 2).  If, for example, the thermal factor in (\ref{ST}) is
replaced by that of free streaming ($\sim\exp(-m_\perp/T)$, see, e.g.,
Refs. \cite{Baym}), then the resulting rapidity fluctuations would
only be the second term in (\ref{seta}). This does not produce the
observed y and $m_\perp$ dependence which indicates that some
thermalization must have taken place in high energy nuclear collisions
before freeze-out.

The source radii above can be Lorenzt boosted longitudinally to any
frame by simply boosting the rapidity $Y$. In the longitudinal
center-of-mass system, defined such that every pair is boosted
to its c.m.s. system, the radii are obtained from the above by setting
$Y=0$. Thus the asymmetric term $R_{ol}$ vanish in Eq. (\ref{Rolb}) to
quadratic order but as pointed out in \cite{Heinz} terms of fourth
order are non-vanishing.

Most resonances produced in relativistic nuclear collisions cannot
decay into two positively or negatively charged pions.  Also, a
single string cannot produce two particles with the same nonvanishing
charge next to each other and an anticorrelation appears.
These anti-correlations in a single resonance or string are, however,
completely washed out by the abundance of resonances in relativistic
nuclear collisions. 

\section{Transverse Expansion}

Besides longitudinal expansion the source may also expand
transversally. Transverse flow has been studied
with renewed interest \cite{Csorgo,Heinz} since it decreases
the source sizes $R_o$ and $R_s$ with increasing $m_\perp$, an effect
that has recently been seen experimentally \cite{NA44}. The analysis
in \cite{Csorgo,Heinz} will here be extended to include resonances.

A source expanding transversally with velocity $v(x)$ has flow velocity
\begin{eqnarray}
  u=\gamma(v)(\cosh(\eta),v_x(x),v_y(x),\sinh(\eta)) \, ,
\end{eqnarray}
which gives
\begin{eqnarray}
   K\cdot u = m_\perp\gamma(v)(\cosh(\eta-Y)-\beta_\perp v_x) \, .
\end{eqnarray}
Assuming the source is expanding as $v_x=vx/\sqrt{\sigma_\perp}$ and
similarly for $v_y$ one finds when $v^2\ll 1$
\begin{eqnarray}
  R_s^2 &=& \sigma_\perp\langle\frac{1+2v\tau_r/\sigma_\perp^{1/2}+
          2v^2\tau_r^2/\sigma_\perp}{1+v^2m_\perp/T_r}\rangle 
                \label{Rsv}\\
  R_o^2 &=& R_s^2+ (\beta_\perp^2+v^2)\langle\tau_r^2\rangle
         +\beta_\perp^2\left[(1+v^2)\sigma(\tau_r)+\sigma(\tau)
         +\langle\tau+(1+v^2)\tau_r\rangle^2\tanh^2Y\sigma(\eta)\right]   \\
  R_l^2 &=&  \sigma(\eta)\left[ \langle\tau^2\rangle
             +2(1+v^2)\langle\tau\tau_r\rangle
              +2(1+v^2)^2\langle\tau_r^2\rangle \right] \cosh^{-2}(Y)    \\
  R_{ol}^2 &=&  - \beta_\perp \sigma(\eta)\langle\tau+(1+v^2)\tau_r\rangle^2 
                \sinh(Y)\cosh^{-2}(Y)  \, .   \label{Rolv}
\end{eqnarray}
One notices the factor $(1+v^2m_\perp/T)^{-1}$ which was found in
earlier analyses \cite{Csorgo,Heinz} and which leads to smaller apparent
sources at large $m_\perp$. Resonances do not change this dependence but
lead to generally larger apparent sources by additional terms of order
$v\tau_r/\sigma_\perp$.

The transverse flow results in smaller transverse momentum slopes or
equivalently larger apparent or effective temperatures, $T_{eff}$.
For a fixed $m_\perp$ slope, the local temperatures are ``red
shifted'' approximately as $T=T_{eff}\sqrt{(1-v)/(1+v)}$
\cite{Schnedermann} as the transverse flow increases.
Consequently, the transverse flow results in smaller source sizes at
large $m_\perp$ not only due to the explicit $v^2$ terms in
Eqs. (\ref{Rsv}-\ref{Rolv}) but also by reducing the temperature.


\begin{figure}
\centerline{\psfig{figure=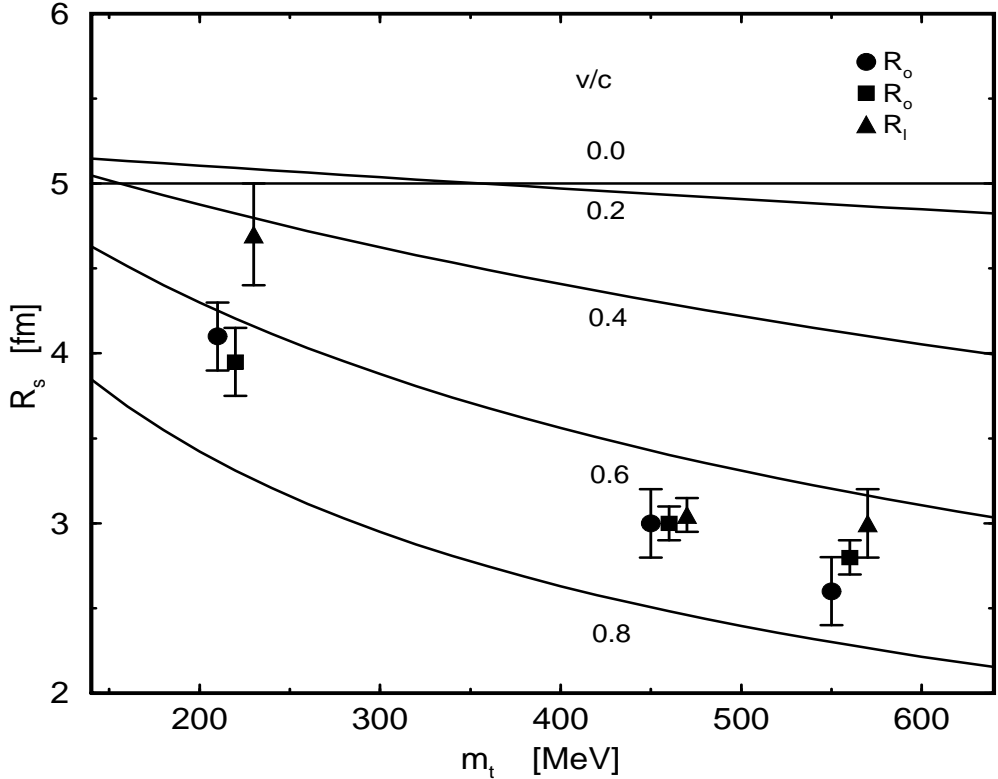,width=16cm,height=12cm,angle=-90}}
\vspace{5mm}
\caption{Effects of transverse flow on the sidewards radius 
as function of transverse mass (see text). Curves correspond from top and
downwards to
$v/c=0,0.2,0.4,0.6,0.8$ . Experimental 200 GeV/A S+Pb data 
\protect{\cite{NA44}} on $R_s$, $R_o$, and $R_l$ are shown (with squares,
circles and diamonds respectively) 
for pions at two different $m_\perp$ and for kaons at $m_\perp=550$MeV. }
\label{fig2}
\end{figure}

In Fig. 2 the $m_\perp$ dependence of the sidewards radius is shown
for $T_{eff}=170$MeV \cite{NA44}, $\sigma_\perp=5$fm, and for various
transverse flow velocities. As seen from Eq. (\ref{Rsv}), the scale is
mainly determined by $\sigma_\perp$ whereas the curvature depends on
the transverse flow velocity. The decreasing size with increasing
$m_\perp$ can be reproduced with flow velocities of order $v/c\simeq
0.6-0.8$.  These transverse velocities are smaller than those obtained
by Cs\"org\H o and L\"orstad. Resonance fractions from RQMD (Table I)
have been employed but using fractions from Fritiof does not change
the dependence on $m_\perp$ by much. The temperature is assumed to
decrease with transverse flow velocity according to the red-shift
formula. Including resonance fractions varying with $m_\perp$ as,
e.g., in RQMD makes only minor differences in the $m_\perp$ dependence
of the source sizes.

\section*{Discussion}

A general description of correlations and source size dependence on
resonances and flow has been given. By expanding to second order in
relative momenta the source radii were extracted analytically and a
nontrivial interplay between the effect of resonances and flow was
found. The source sizes receives contributions from fluctuations in
transverse spatial directions, source life time, rapidity and
resonances life times.  The long lived resonances lead to a reduction
in the correlation function and can explain most of the reduction in
the correlations function measured experimentally,
$\lambda^{\pi\pi}\simeq0.5-0.6$, when the experimental cuts are
included.  Due to strangeness enhancement in RQMD of in particular the
long lived $K^0_S$, it gives a smaller $\lambda$ in better agreement with
experiment than Fritiof.  
However, the similar values for $\lambda^{\pi\pi}$ in {\it pp} and $p\bar{p}$
($\lambda\simeq 0.40$ \cite{pp}\footnote{The {\it pp} and $p\bar{p}$
data are, however, at higher collision energy, $\sqrt{s}=63$GeV, and
at larger relative momentum, $q>50$MeV/c.}) and {\it p}-nucleus
($\lambda^{pA}\simeq 0.41$ \cite{NA44}) and nuclear collisions do
not indicate that rescattering, thermalization, strangeness
enhancement or other collective effects in nuclear collisions are
affecting the resonance production and $\lambda$ significantly. 
There is no indication of the $\omega$-resonance in the
data, which is otherwise abundant in most models. From measurements of the $KK$
correlation functions one finds $\lambda^{KK}=0.83\pm0.08$ \cite{NA44}), which
is close to $\lambda_{cut}$, and therefore no long lived resonances
decaying into kaons are required. This is predicted in the models
discussed above where kaons mainly receive feed-down from the
relatively short-lived $K^*$.

Even though resonances and cuts can explain most of the reduction in
$\lambda^{\pi\pi}$ found in nuclear collisions, 
$\lambda^{\pi\pi}\simeq 0.5-0.6$ \cite{NA44,WA80,AGS}, 
it does not exclude a partially coherent pion source.
However, the  kaon sources
measured, $\lambda^{KK}/\lambda_{cut}\sim1$, are apparently incoherent.  
Furthermore, if particles are produced
coherently within a coherence length, $\xi_{coh}$, in space and time,
the reduction in the correlation function would be of order
$\lambda\sim 1-(\xi_{coh}/R)^4$.  The similar $\lambda$ values found in
{\it pp, p}$\bar{p}$, {\it p}-nucleus and 
nucleus-nucleus collisions would then require that the correlation
length scales approximately with the source size. 

Other effects, that might explain the low $\lambda^{\pi\pi}$, are
final state interactions and Coulomb repulsion. 
Strong interactions are generally found to be insignificant in
nuclear collisions due to their short interaction range as compared to 
nuclear length scales.
Coulomb repulsion reduce the
correlation function significantly at small relative momenta, which is
usually corrected for by the Gamov factor
$\Gamma=\tilde{\eta}/(e^{\tilde{\eta}}-1)$, where $\tilde{\eta}=2\pi
me^2/Q_{inv}\simeq 6{\rm MeV}/Q_{inv}$ for pions.  Coulomb screening
effects are only of order a few percent \cite{Anchiskin} but as
they reduce the Coulomb interactions, they lead to a {\it smaller}
$\lambda$. 

Experimental data seem to indicate both longitudinal and transverse flow.
Longitudinal flow as in the Bjorken model seems to agree with the
rapidity dependence of the longitudinal source radius.  The decreasing
transverse source sizes with increasing transverse mass of the
particles can be explained in a simple model with transverse flow of
order $v=0.5-0.8c$ at the surface when resonances are included.  This
assumes an apparent blue-shifted temperature that is fixed by the
transverse momentum slopes of particle spectra.  The source sizes have
been measured for two values for the pion transverse mass and one kaon
transverse mass.  The smaller kaon radii can thus be explained solely
by transverse flow. The conventional explanation is that the smaller
kaon scattering cross section results in earlier freeze-out at smaller
radii than for pions. However, such a cascade picture leads to an
extended particle emission time typically of the same order as the
expansion time, $\sigma(\tau)\sim\langle\tau\rangle$, and not a sharp
freeze-out in time as is indicated by the similarity of the sidewards
and outwards radii.  Also the freeze-out in hydrodynamic models at a
constant temperature takes about as long as the expansion time.  This
is in contradiction with the very similar outwards and sidewards
transverse radii measured in particular when the resonance life-times
are taken into account.

Clearly, more data is crucial in order to describe the sources and
determine their size, lifetime, longitudinal and
transverse expansion, so that we can discriminate between these models. In
particular, it would be most useful if one could measure the feed-down from
resonances independently. One would then be able to predict
the non-gaussian correlation function (see Fig. 1) and 
separate the effect of resonances from coherence, final state interactions etc.
Only when resonances are properly accounted for and their effect on
the correlation function,
$\lambda$, the source radii and life time are understood and corrected for, can
we extract the ``bare'' source sizes and the flow reminiscent of 
the earlier phase of hot and dense matter created in high energy nuclear 
collisions.

\section*{Acknowledgements}

Discussions with Hans B\o ggild, Bengt L\"orstad, Heinz Sorge, John
Sullivan and Axel Vischer are gratefully acknowledged.

\vspace{2cm}
\begin{flushright}Preprint Nordita-96/15 N\end{flushright}
\begin{flushright}hep-ph/9602431\end{flushright}


\end{document}